\documentclass{ifacconf}
\usepackage{graphicx}      % include this line if your document contains figures
\usepackage{natbib}        % required for bibliography
\usepackage{xcolor} % Ensure this is in the preamble
\usepackage{circuitikz}
\usepackage{graphicx} % in preamble
\usepackage{amsmath,amssymb} %‌,amsthm
  
\DeclareMathSizes{10}{10}{6}{6}
\usepackage{appendix}

\begin{document}
\begin{frontmatter}

\title{Equilibrium points and stability of synchronous machine systems} 

\author[First]{Maryam Khodabakhshloo} 
\author[second]{Elizabeth L. Ratnam} 
\author[First]{Ian R. Petersen}

\address[First]{School of Engineering, The Australian National University,
Canberra, Australia (e-mail: maryam.khodabakhshloo, ian.petersen)@anu.edu.au.}
\address[second]{Department of Electrical and Computer
Systems Engineering, Monash University, Melbourne, Australia
(e-mail: liz.ratnam@monash.edu).}

% \author[First]{Maryam Khodabakhshloo} 
% \author[First]{Elizabeth L. Ratnam} 
% \author[First]{Ian R. Petersen} 

% \address[First]{The School of Engineering, The Australian National University,
% Canberra, Australia (e-mail: maryam.khodabakhshloo, elizabeth.ratnam, ian.petersen)@anu.edu.au.}
\begin{abstract}
This paper investigates equilibrium points and stability in two synchronous machine configurations: (i) a single generator with an impedance load and (ii) two interconnected machines with co-located loads. We consider both $abc$ and $dq$ reference frames to show that the equilibrium condition reduces to a cubic polynomial in the single-machine case and to an 18th-degree polynomial in the two-machine case. For the single-machine system, Lyapunov stability analysis and linearization based stability analysis are carried out.
 For the two-machine system, local stability is assessed through linearization and eigenvalue analysis. Illustrative examples confirm the existence of multiple equilibria and illustrate the impact of parameter variation on stability. Our results provide insight into the stability of synchronous machine systems.
\end{abstract}


\end{frontmatter}
%===============================================================================
\section{Introduction}
\label{sec:intro}
Rapid growth in grid-connected renewable energy zones and behind-the-meter, inverter-based resources are transforming grid dynamics. Accordingly, the Stable operation of modern power systems is increasingly difficult to ensure.  Historically, power systems have been dominated by synchronous machines (SMs), with stability assessments limited to transmission-level models or otherwise, steady-state assumptions. While these approaches provide useful insights, it can be difficult to capture the fast and nonlinear behaviors that emerge during large disturbances under renewable-dominated power grids \cite{anderson2008power, machowski2020power}. This limitation highlights the need for analytical tools that can deliver fast, reliable, and physically interpretable stability assessments for future low-inertia grids, where power-electronics dynamics play a central role.  

Synchronous machines have historically been the foundation of power system operation and are commonly modeled by reduced-order representations, such as the classical second-order swing equation \cite{machowski2020power, schiffer2016survey}. These models greatly simplify analysis, but at the cost of neglecting the electrical transients associated with stator dynamics and network impedance effects. Moreover, they rely on linearizations around nominal operating points, which obscure important nonlinear phenomena. As a result, such models can misrepresent stability margins when synchronous generators interact with converter-interfaced renewable sources or with frequency-dependent loads \cite{zhou2008improved,bergen2007structure}. Consequently, there has been renewed interest in high-fidelity electromechanical models that preserve the nonlinearities necessary for rigorous analysis \cite{leonov2006phase,zhong2010synchronverters}.  

To assess the stability of synchronous machine systems, many authors have considered the \textit{dq}-transformation, which was developed for instantaneous power analysis of three-phase systems \cite{hirofumi2007instantaneous,teodorescu2011grid}. The \textit{dq}-transformation simplifies the equations representing electromechanical interactions, and has inspired the development of converter-based control that
 mimics the behavior of conventional SMs through power-electronics \cite{zhong2010synchronverters}. However, the literature in \cite{du2019power,gholami2020fast,ma2021generalized,qiu2020swing}
 % \cite{liu2023coupling} 
 applies linearization techniques to assess the local stability of power systems.
 In contrast, the authors in  \cite{oh2019analytical,trip2016internal} construct Lyapunov-based energy functions for simplified reduced-order models. 
 Accordingly, such methods are restricted to small-signal analysis or to cases where suitable Lyapunov functions can be explicitly derived.  
 
In this paper, we consider a detailed equilibrium and stability analysis for two fundamental synchronous machine system configurations.
 First, we derive exact equilibrium conditions for; (i) a single synchronous machine with an impedance load, which reduces to a cubic polynomial in the equilibrium angular frequency, and (ii) two interconnected synchronous machines with local loads, which lead to an 18th-degree polynomial. 
Next, for the single-machine case, we apply both Lyapunov analysis—using an energy-based candidate Lyapanov function to obtain sufficient conditions for global stability—and linearization to characterize local stability around the equilibria. For the two-machine case, we focus on linearization and eigenvalue analysis to assess local stability.  Illustrative examples validate the analytical results and reveal a parameter-dependent multiplicity of equilibria. 

The remainder of this paper is organized as follows. 
Section~\ref{sec:modeling} introduces the synchronous machine systems and applies the \textit{dq}-transformation. 
Section~\ref{sec:equilibrium} derives the equilibrium conditions. 
In Section~\ref{sec:stability}, we consider conditions for the stability of synchronous machine systems together with illustrative examples. 
Section~\ref{sec:conclusion} provides concluding remarks.

\section{Synchronous Machine Systems}
\label{sec:modeling}
In this section, we present the electromechanical equations for a synchronous machine (SM). Specifically, we introduce the \textit{dq}-coordinate transformation, and its application for two circuit configurations: 1) an SM connected to an impedance load; 2) two SMs interconnected via a transmission line, each with a co-located impedance load. For both configurations, the electromechanical equations are developed in the \textit{abc} and \textit{dq} reference frames.
\subsection{Synchronous Machine}\noindent
%Following the established electromechanical modeling approaches \cite{grainger1999power,leonov2006phase,zhong2010synchronverters,schiffer2016survey,barabanov2017conditions}, the SM is modeled under several simplifying assumptions. 
We consider a synchronous machine that generates electricity and incorporates a stator and a rotor. 
The stator windings are where alternating current (AC) flows, producing an alternating magnetic field at the same frequency. The rotor is cylindrical, and we consider a synchronous machine with one pole pair per phase.
We neglect effects such as damper windings, magnetic saturation, and eddy currents. 
We assume the rotor current $ i_f$ is constant. The stator is excited by a balanced three-phase signal in a star-connected configuration without a neutral wire.
The stator windings are sinusoidally distributed and symmetric, resulting in a smoothly rotating magnetic field. The generator’s reference direction is chosen so that the current flowing out of the generator is considered positive.

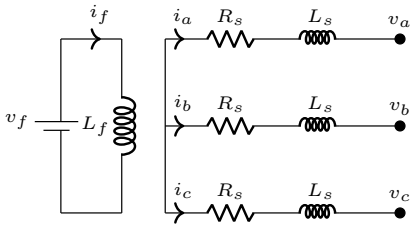
\begin{figure}[htbp]
    \centering
    \scalebox{1.15}{ % <-- scaling starts here
    \begin{circuitikz}
    \tiny
        % Sinusoidal voltage source (Generator)
        % \draw (-3,0) to[sinusoidal voltage source, l_=$ SG $] (0,0);
        % \draw[-stealth, red, thick] (-2.5,-.25) arc[start angle=-80,end angle=150,radius=0.25];
        % \node at (-2.5,-0.5) {$(T,\omega)$};
        % \draw (0,-2) -- (5.1,-2);
        \draw [color=black](-.5,1) to[L,bipoles/length=30pt] (-.5,-1) ;
        \draw [color=black](0,0) to[R, bipoles/length=18pt] (1.5,0) to[L, bipoles/length=18pt] (2,0);
        \draw [color=black](0,1) to[R, bipoles/length=18pt] (1.5,1) to[L, bipoles/length=18pt] (2,1);
        \draw [color=black](0,-1)to[R,bipoles/length=18pt] (1.5,-1) to[L, bipoles/length=18pt] (2,-1);
 
        %%%%%%%%%%%%%%%%%%%%%%%%%%%%%%
        \draw[fill=black] (2.7,1) circle (1.62pt);
        \draw[fill=black] (2.7,0) circle (1.62pt);
        \draw[fill=black] (2.7,-1) circle (1.62pt);
        
        %%%%%%%%%%%%%%%%%%%%%%%%%%%%%%
        \draw (-1.2,1) -- (-.5,1);
        \draw (-1.2,-1) -- (-.5,-1);
        %%%%%%%%%%%%%%%%%%%%%%%%%%%%%%
        \draw (0,1) -- (0,-1);
        %%%%%%%%%%%%%%%%%%%%%%%%%%%%
        \draw (2,1) -- (2.7,1);
        \draw (2,0) -- (2.7,0);
        \draw (2,-1) -- (2.7,-1);
        %%%%%%%%%%%%%%%%%%%%%%%%%%%%%%%%%
        \draw (-1.2,1) -- (-1.2,.05);
        \draw (-1.2,-1) -- (-1.2,-.05);

        %%%%%%%%%%%%%%%%%%%%%%%%%%%%%
        \draw (-1.5,.05) -- (-1,.05);
        \draw (-1.4,-.05) -- (-1.1,-.05);

        \node at (-1.7,.05) {$v_f$};       
        
        % \node at (-1.7,.051) {$+$};       
        % \node at (-1.7,-.051) {$-$};    
        %%%%%%%%%%%  L_s %%%%%%%%%%%%%
        \node at (.75,1.25) {$R_s$};
        \node at (.75,.25) {$R_s$};
        \node at (.75,-.75) {$R_s$};  
        %%%%%%%%%%%% L_f %%%%%%%%%%%%%
        \node at (-.8,0) {$L_f$};  
        % \node at (-2,0) {$DC$};  

        %%%%%%%%%%%  R_s %%%%%%%%%%%%%
        \node at (1.79,1.25) {$L_s$};
        \node at (1.79,.25) {$L_s$};
        \node at (1.79,-.75) {$L_s$};
        %%%%%%%%%%% V_abc %%%%%%%%%%%%
        \node at  (2.7,1.2)   {$v_a$};
        \node at (2.7,.2)   {$v_b$};
        \node at  (2.7,-.8)   {$v_c$};    
        %%%%%%%%%%%%%%%%%%%%%%%%%%%%%%
        \draw[->, thick] (-.75,1) -- (-.74,1) node[midway, above, yshift=2pt]{$i_{f}$};
        %%%%%%%%%%%%%%%%%%%%%%%%%%%%%%
        \draw[->, thick] (.2,1) -- (.22,1) node[midway, above, yshift=2pt]{$i_{a}$};
        \draw[->, thick] (.2,0) -- (.22,0) node[midway, above, yshift=2pt]{$i_{b}$};
        \draw[->, thick] (.2,-1) -- (.22,-1) node[midway, above, yshift=2pt]{$i_{c}$};        
    \end{circuitikz}}
\caption{\centering{Synchronous Machine equivalent circuit.}}
\label{SG_M}
\end{figure}

\noindent
The following notation largely follows that of \cite{zhong2010synchronverters}.
The stator resistance and inductance are denoted by \( R_s \) and \( L_s \), respectively, and \( i_f \) is the constant rotor current (see in Figure \ref{SG_M}). The rotor electrical phase angle, angular velocity and the corresponding stator current; and voltage are denoted by \( \theta \in \mathbb{R} \), \( \omega = \dot{\theta} \in \mathbb{R} \), \( i_{abc} \in \mathbb{R}^3 \), and \( v_{abc} \in \mathbb{R}^3 \), respectively.
\color{black}
\color{black}

\noindent
The electromotive force \( e_{abc} \) induced in the stator by the rotating magnetic field generated by the rotor motion is given by:
\begin{equation}
    e_{abc} = M_f i_f \omega
    \begin{bmatrix}
        \sin(\theta) \\
        \sin\left(\theta - \frac{2\pi}{3}\right) \\
        \sin\left(\theta - \frac{4\pi}{3}\right)
    \end{bmatrix}
\label{backemf}
\end{equation}
where \( M_f \) is the mutual inductance between the stator and the rotor, and
$\theta$ is the rotor angle with respect to the fixed reference angle and \(\omega\) is the rotor angular velocity, also referred to as the frequency.

\noindent
By applying Ohm’s law and the inductor law, the voltages \(v_{abc}\) at each terminal \textit{abc} of the stator satisfy the equations:
\[
    v_{abc}  = -R_s i_{abc}  - L_s \frac{di_{abc}}{dt} + e_{abc}.  
% \label{SGMV}
\]

\noindent
Thus, the electrical equations describing the dynamics of the three-phase stator current are:
\[
    L_s \frac{di_{abc}}{dt} = -R_s i_{abc} + e_{abc} - v_{abc}.
\]

\noindent
The mechanical dynamics of the rotor are described by Newton’s second law, equating the net electromagnetic and mechanical torques to the inertial torque. Specifically,
\[
J \dot{\omega} = -D\omega + T_m - T_e,
\]
where \(J\) is the rotor moment of inertia, \(D\) is a damping coefficient and \(T_m\) is the input mechanical torque. The electrical torque \( T_e \), with the assumption of constant field current, is derived as in \cite{zhong2010synchronverters,caliskan2014compositional} as
\begin{equation}
T_e = \dot{\theta}^{-1} \langle  e_{abc}, i_{abc} \rangle = \omega^{-1} e_{abc}^\top i_{abc}.
\label{electricaltorque}
\end{equation}

\noindent
The next section begins with a brief overview of the \textit{dq} framework. Then we introduce two synchronous machine system configurations and transform the electromechanical equations from the \textit{abc} frame into the \textit{dq} frame, simplifying further dynamical analysis.
\noindent
\subsection{\textit{dq}-Coordinate Transformation}

The \( dq \)-coordinate, also known as the rotating reference frame~\cite{schiffer2016survey}, is a mathematical transformation that maps three-phase \( abc \) quantities into two orthogonal components. This transformation simplifies the analysis of balanced electrical systems and is widely employed in power system control applications. 
The \( dq \)-transformation is given by,
\[
U(\eta) = \sqrt{\frac{2}{3}} 
\begin{bmatrix}
    \cos(\eta) & \cos\left(\eta - \frac{2\pi}{3}\right) & \cos\left(\eta - \frac{4\pi}{3}\right) \\
    \sin(\eta) & \sin\left(\eta - \frac{2\pi}{3}\right) & \sin\left(\eta - \frac{4\pi}{3}\right) 
    % \\
    % \frac{\sqrt{2}}{2} &\frac{\sqrt{2}}{2} &\frac{\sqrt{2}}{2}
\end{bmatrix}.
\]  
where  \(\eta\) is the reference angle.
We use the following properties of the \( dq \)-transformation in subsequent sections:
\begin{enumerate}
    \renewcommand{\labelenumi}{(\roman{enumi})}
    \item %Orthonormality; \( \qquad U(\eta)=U(\eta)^{-1}. \)
    % Row-orthonormality;\( U(\eta)U(\eta)^\top=I. \)    
       % Row-orthonormality;   \( U(\eta)U(\eta)^\top=I. \) implies  
       $$ \quad i_{dq}=U(\eta)i_{abc} \quad , \quad  i_{abc}=U(\eta)^\top i_{dq}.$$
       
    % \item Composition rules;
    % \begin{equation}
    %     % U(\eta_1) U(\eta_2) = U(\eta_1 + \eta_2),
    %      U(\eta_1) U(\eta_2)^\top = U(\eta_1 - \eta_2).
    % \label{composition_rule}
    % \end{equation}
    \item time-derivative relationship;
    \begin{equation}
    \frac{d i_{dq}}{dt} = \dot{\eta} \begin{bmatrix}
        0&-1\\
        1&0
    \end{bmatrix} U(\eta) \, i_{abc} + U(\eta) \frac{d i_{abc}}{dt}.
    \label{abcmoshtagh}
    \end{equation}
\end{enumerate}
\subsection{Synchronous generator with an impedance load}

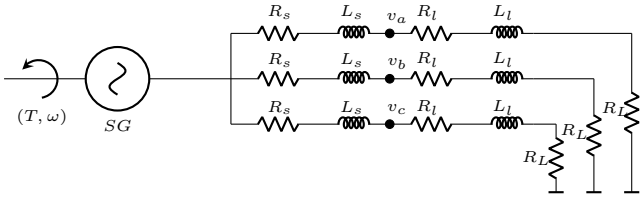
\begin{figure}[h!t]
    \centering
    \scalebox{1}{ % <-- scaling starts here
    \begin{circuitikz}
    \tiny
        % Sinusoidal voltage source (Generator)
        \draw (-3,0) to[sinusoidal voltage source, l_=$ SG $] (0,0);
        \draw[-stealth, black, thick] (-2.5,-.25) arc[start angle=-80,end angle=150,radius=0.25];
        \node at (-2.5,-0.5) {$(T,\omega)$};
        % \draw (0,-2) -- (5.1,-2);
        \draw [color=black](0,0) to[R, bipoles/length=18pt] (1.25,0) to[L, bipoles/length=18pt] (2,0) to[R, bipoles/length=18pt] (3.3,0)to[L, bipoles/length=18pt] (4,0); % 
        \draw [color=black](0,.6) to[R, bipoles/length=18pt] (1.25,.6) to[L, bipoles/length=18pt] (2,.6) to[R, bipoles/length=18pt] (3.3,.6)to[L, bipoles/length=18pt] (4,.6); %     
        \draw [color=black](0,-.6) to[R, bipoles/length=18pt] (1.25,-.6) to[L, bipoles/length=18pt] (2,-.6) to[R, bipoles/length=18pt] (3.3,-.6)to[L, bipoles/length=18pt] (4,-.6); % 
            
        % \draw (0,0) -- (.35,0);
        
        \draw [color=black] (5.3,.6) to[R, bipoles/length=18pt] (5.3,-1.5);
        \draw [color=black] (4.8,0) to[R, bipoles/length=18pt] (4.8,-1.5);
        \draw [color=black] (4.3,-.6) to[R, bipoles/length=18pt] (4.3,-1.5);

        %%%%%%%%%%%%%%%%%%%%%%%%%%%%%%%%%%%%%%%%%%%%%%%%%%%%
        \draw[fill=black] (2.1,.6) circle (1.62pt);
        \draw[fill=black] (2.1,0) circle (1.62pt);
        \draw[fill=black] (2.1,-.6) circle (1.62pt);

        %%%%%%%%%%%%%%%%%%%%%%%%%%%%%%%%%
        
        %%%%%%%%%%%%%%%%%%%%%%%%%%%%%%%%%
        \draw (0,.6) -- (0,-.6);
        %%%%%%%%%%%%%%%%%%%%%%%%%%%%%%%%%5
        \draw (4,.6) -- (5.3,.6);
        \draw (4,0) -- (4.8,0);
        \draw (4,-.6) -- (4.3,-.6);
        
        %%%%%%%%%%%%%%%%%%%%%%%%%%%%%%%%%
        \draw[thick] (5.2,-1.5) -- (5.4,-1.5);
        \draw[thick] (4.7,-1.5) -- (4.9,-1.5);
        \draw[thick](4.2,-1.5) -- (4.4,-1.5);
        
        %%%%%%%%%%%  R_L %%%%%%%%%%%%%%%%
        \node at (5.09,-.4) {$R_L$};
        \node at (4.55,-.7) {$R_L$};
        \node at (4.04,-1.05) {$R_L$};

        %%%%%%%%%%%  R_s %%%%%%%%%%%%%%%%
        \node at (.65,.88) {$R_s$};
        \node at (.65,.25) {$R_s$};
        \node at (.65,-.38) {$R_s$};
        
        %%%%%%%%%%%  L_s %%%%%%%%%%%%%%%%
        \node at (1.6,.88) {$L_s$};
        \node at (1.6,.25) {$L_s$};
        \node at (1.6,-.38) {$L_s$};
        
        %%%%%%%%%%%  R %%%%%%%%%%%%%%%%
        \node at (2.6,.88) {$R_l$};
        \node at (2.6,.25) {$R_l$};
        \node at (2.6,-.38) {$R_l$};
        
        %%%%%%%%%%%  L %%%%%%%%%%%%%%%%
        \node at (3.6,.88) {$L_l$};
        \node at (3.6,.25) {$L_l$};
        \node at (3.6,-.38) {$L_l$};
        
        %%%%%%%%%%%%V_abc%%%%%%%%%%%%%%%
        \color{black}
        \node at  (2.2,.8)   {$v_a$};
        \node at (2.2,.2)   {$v_b$};
        \node at  (2.2,-.4)   {$v_c$}; 
  
    \end{circuitikz}}
\caption{SG connected to an impedance load via a transmission line.}
\label{SG_T_R}
\end{figure}
\noindent
In Figure~\ref{SG_T_R}, we consider a synchronous generator (SG) connected through a transmission line to an impedance load.  
The circuit consists of the stator resistance $R_s$ and inductance $L_s$, the line resistance $R_l$ and inductance $L_l$, and the load resistance $R_L$.  
Since the inductances and resistances are connected in series, we define equivalent parameter values by  
$R = R_s + R_l + R_L$ and $L = L_s + L_l$. Figure~\ref{SG_T_R} presents the electromechanical system, where the three-phase $abc$ frame, is expressed by equations 
\begin{equation}
    \begin{aligned}
        \dot{\theta} &= \omega, \\
        J \dot{\omega} &= -D\omega - \omega^{-1} e_{abc}^\top i_{abc} + T_m, \\
        L \frac{di_{abc}}{dt}  &= -R i_{abc} + e_{abc}.
    \end{aligned}
    \label{electromechanical_modeling_abc}
\end{equation}
\noindent
Next, we transform the \(abc\) model in \eqref{electromechanical_modeling_abc} into the \(dq\) reference frame. We choose the transformation angle \(\eta\) to be equal to the rotor angle,  \(\eta = \theta\). Accordingly, using the relation \(i_{abc} = U^\top(\eta) i_{dq}\), we substitute 
\(i_{abc} \)
 into the mechanical dynamics in \eqref{electromechanical_modeling_abc} to obtain the dynamics in the 
\(dq\)-frame:
% transformation, the mechanical dynamic in \eqref{electromechanical_modeling_abc} is rewritten by substituting \(i_{abc}\),
$$
 \omega^{-1} e_{abc}^\top i_{abc}= \omega^{-1} e_{abc}^\top U^\top(\eta) i_{dq}=\sqrt{\frac{3}{2}}
 M_fi_f\begin{bmatrix}
     0\\
     1
 \end{bmatrix}i_{dq}.
$$
Also, the electrical equation is premultiplied by \(U(\eta)\) and using (\ref{abcmoshtagh}), results in the following \(dq\)-frame model:
\begin{equation}
\begin{aligned}
    \dot{\theta} &= \omega, \\
    J \dot{\omega} &= -D \omega - b i_q + T_m, \\
    L \dot{i}_d &= -R i_d - L \omega i_q, \\
    L \dot{i}_q &=  L \omega i_d - R i_q+b \omega,
\end{aligned}
\label{electromechanical_modeling_dq}
\end{equation}
where \(b=\sqrt{\frac{3}{2}}M_fi_f.\)
\subsection{Two SMs Interconnected via a Transmission Line with co-located Resistive Loads}
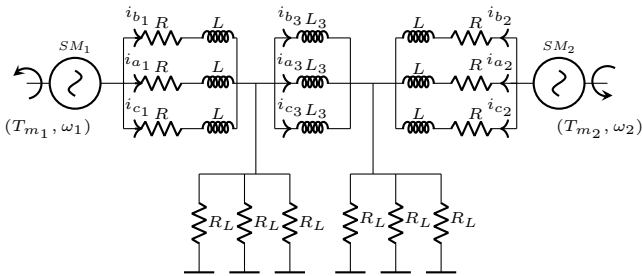
\begin{figure}[htbp]
    \raggedright % Align content to the left
    \scalebox{1}{ % scaling starts here
    \begin{circuitikz}
    \tiny
        % Sinusoidal voltage sources (Generators) in-line
        \draw[scale=.8, transform shape] (0,0) to[sinusoidal voltage source, l_=$ SM_1 $] (-1.6,0);
        \draw[scale=.8, transform shape]  (8,0) to[sinusoidal voltage source, l_=$ SM_2 $] (6.4,0);
        %%%%%%%%%%%%%%%%%%%%%%%%%%%%%%%%%%%%%%%%%%
        % \draw (-2,0) -- (-1.5,0);
        \draw (0,0) -- (0,0);
        % Torque and speed arrows for SGs
        \draw[-stealth, black, thick] (-1.25,-.2) arc[start angle=-80,end angle=150,radius=0.2];
        \node at (-1,-0.6) {\tiny$(T_{m_1},\omega_1)$};
        \draw[-stealth, black, thick] (6.5,.2) arc[start angle=60,end angle=290,radius=0.2];
        \node at (6.3,-0.6) {\tiny$(T_{m_2},\omega_2)$};

        %%%%%%%%%%%%%%%%%%%%%%%%%%%%%%%%%%%%%%%%%%%%%%%%%%%%
        % Left-side R and L
        \draw [color=black](0,0) to[R, bipoles/length=18pt](1,0) to[L, bipoles/length=18pt](1.5,0)(2,0) to[L, bipoles/length=18pt](3,0); 
        \draw [color=black](0,.6)to[R, bipoles/length=18pt](1,.6)to[L, bipoles/length=18pt](1.5,.6)(2,.6)to[L, bipoles/length=18pt](3,.6); 
        \draw [color=black](0,-.6)to[R,bipoles/length=18pt](1,-.6)to[L,bipoles/length=18pt](1.5,-.6)(2,-.6)to[L,bipoles/length=18pt](3,-.6);
        %%%%%%%%%%%%%%%%%%%%%%%%%%%%%%%%%%%%%%%%%%%%%%%%%%%%
        % Right-side R and L
        \draw  [color=black] (3.6,0)to[L, bipoles/length=18pt] (4.2,0)to[R, bipoles/length=18pt](5,0)--(5.2,0);     
        \draw  [color=black] (3.6,.6)to[L, bipoles/length=18pt] (4.2,.6)to[R, bipoles/length=18pt](5,.6)--(5.2,.6);
        \draw  [color=black] (3.6,-.6)to[L, bipoles/length=18pt] (4.2,-.6)to[R, bipoles/length=18pt](5,-.6)--(5.2,-.6);
        
        % Left R_L bank
        \draw [color=black] (1,-1.2) to[R, bipoles/length=18pt] (1,-2.5);
        \draw [color=black] (1.6,-1.2) to[R, bipoles/length=18pt] (1.6,-2.5);
        \draw [color=black] (2.2,-1.2) to[R, bipoles/length=18pt] (2.2,-2.5);
        \draw[thick] (.8,-2.5) -- (1.2,-2.5);
        \draw[thick] (1.4,-2.5) -- (1.8,-2.5);
        \draw[thick](2,-2.5) -- (2.4,-2.5);
        \node at (1.3,-1.85) {$R_L$};
        \node at (1.9,-1.85) {$R_L$};
        \node at (2.5,-1.85) {$R_L$};

        % Right R_L bank
        \draw [color=black] (3,-1.2) to[R, bipoles/length=18pt] (3,-2.5);
        \draw [color=black] (3.6,-1.2) to[R, bipoles/length=18pt] (3.6,-2.5);
        \draw [color=black] (4.2,-1.2) to[R, bipoles/length=18pt] (4.2,-2.5);
        \draw[thick] (2.8,-2.5) -- (3.2,-2.5);
        \draw[thick] (3.4,-2.5) -- (3.8,-2.5);
        \draw[thick](4,-2.5) -- (4.4,-2.5);
        \node at (3.3,-1.8) {$R_L$};
        \node at (3.9,-1.8) {$R_L$};
        \node at (4.5,-1.8) {$R_L$};
        
        % Vertical connections
                \draw (1.5,.6) -- (1.5,-.6);
                \draw (1.5,0) -- (2,0);
                \draw (1.75,0) -- (1.75,-1.2);
                \draw (1,-1.2) -- (2.2,-1.2);
                \draw (0,.6) -- (0,-.6);
                \draw (3,.6) -- (3,-.6);
                \draw (2,.6) -- (2,-.6);
                \draw (3.3,0)--(3.3,-1.2);
                \draw (3.6,.6)--(3.6,-.6);
                \draw (5.2,.6)--(5.2,-.6);
                \draw (3,0)--(3.6,0);
                \draw (3,-1.2)--(4.2,-1.2);
        %%%%%%%%%%%  R_s %%%%%%%%%%%%%%%%
        \node at (.5,.8) {$R$};
        \node at (.5,.215) {$R$};
        \node at (.5,-.4) {$R$};
        
        %%%%%%%%%%%  L_s %%%%%%%%%%%%%%%%
        \node at (1.25,.8) {$L$};
        \node at (1.25,.215) {$L$};
        \node at (1.25,-.4) {$L$};
        
        %%%%%%%%%%%  L_3 %%%%%%%%%%%%%%%%
        \node at (2.55,.83) {$L_3$};
        \node at (2.55,.23) {$L_3$};
        \node at (2.55,-.37) {$L_3$};
        
        %%%%%%%%%%%  R_s (right) %%%%%%%%
        \node at (4.65,.8) {$R$};
        \node at (4.65,.25) {$R$};
        \node at (4.65,-.4) {$R$};
        
        %%%%%%%%%%%  L_s (right) %%%%%%%%
        \node at (3.9,.8) {$L$};
        \node at (3.9,.2) {$L$};
        \node at (3.9,-.4) {$L$};

        %%%%%%%%%%% Currents %%%%%%%%%%%%
        \draw[->, thick] (.15,0) -- (.25,0) node[midway, above, yshift=2pt]{$i_{a_1}$};
        \draw[->, thick] (.15,.6) -- (.25,.6) node[midway, above, yshift=2pt]{$i_{b_1}$};
        \draw[->, thick] (.15,-.6) -- (.25,-.6) node[midway, above, yshift=2pt]{$i_{c_1}$};

        \draw[->, thick] (2.21,0) -- (2.22,0) node[midway, above, yshift=2pt]{$i_{a_3}$};
        \draw[->, thick] (2.21,.6) -- (2.22,.6) node[midway, above, yshift=2pt]{$i_{b_3}$};
        \draw[->, thick] (2.21,-.6) -- (2.22,-.6) node[midway, above, yshift=2pt]{$i_{c_3}$};
        
        \draw[->, thick] (5,0) -- (4.98,0) node[midway, above, yshift=2pt]{$i_{a_2}$};
        \draw[->, thick] (5,.6) -- (4.98,.6) node[midway, above, yshift=2pt]{$i_{b_2}$};
        \draw[->, thick] (5,-.6) -- (4.98,-.6) node[midway, above, yshift=2pt]{$i_{c_2}$};

    \end{circuitikz}}
    \caption{Two SMs supplying resistive loads and connected via a transmission line.}
    \label{TSG_CT_SL}
\end{figure}

\noindent
In Figure~\ref{TSG_CT_SL}, we present two identical synchronous machines (SMs) that are connected through a transmission line, and each supplies power to resistive loads.

\noindent
We assume all electrical parameters are balanced. 
The stator of each synchronous machine is identical, each characterized by a stator resistance $R  \ge 0$ and stator inductance $L  \ge 0$. 
The transmission line is represented by an inductance $L_3$, and the load is represented by a resistance $R_L > 0$. 
For each SM, the damping coefficient and moment of inertia are denoted by $D >0, J >0 $, respectively.
The mechanical torque input, the rotor angle, electrical angular velocity, and generated current of the \( j \)-th SM is denoted by $T_{m_j} \ge 0$ , \( \theta_j(t) \in \mathbb{R} \), \( \omega_j(t) \in \mathbb{R} \), and \( i_{abc,j}(t) \in \mathbb{R}^3 \), respectively, where $j = 1, 2$ represents the indices of the machines. The electromotive forces \( e_{abc_1}, e_{abc_2} \) are defined in Equation \eqref{backemf}, and the electrical torques \( T_{e_1}, T_{e_2} \) are defined in Equation \eqref{electricaltorque}.
\noindent
The electromechanical dynamics of the system shown in Figure~\ref{TSG_CT_SL}, expressed in \( abc \)-coordinates, are given by:
\begin{equation}
\begin{array}{l}
\dot{\theta}_1 = \omega_1, \\
\dot{\theta}_2 = \omega_2, \\
\\
J \dot{\omega}_1 = -D \omega_1 - \omega_1^{-1} e_{abc_1}^\top i_{abc_1} + T_{m_1}, \\
J \dot{\omega}_2 = -D \omega_2 - \omega_2^{-1} e_{abc_2}^\top i_{abc_2} + T_{m_2}, \\
\\
L \frac{d i_{abc_1}}{d t} = -(R + R_L) i_{abc_1} + e_{abc_1} + R_L i_{abc_3}, \\
L \frac{d i_{abc_2}}{d t} = -(R + R_L) i_{abc_2} + e_{abc_2} - R_L i_{abc_3}, \\
L_3 \frac{d i_{abc_3}}{d t} = -2 R_L i_{abc_3} + R_L (i_{abc_1} - i_{abc_2}).
\end{array}
\label{TSG_CT_SL3phasemodel}
\end{equation}

% \subsubsection{\textit{dq0} Modeling}
\noindent
Next, we transform the system equations in (\ref{TSG_CT_SL3phasemodel}) into the \textit{dq} reference frame. The transformation angle \(\eta\) is taken to be the average of the two SMs rotor angles, defined by
\begin{equation}
    \eta = \frac{1}{2}(\theta_1 + \theta_2).
\label{transangleTSGTRL}
\end{equation}
Accordingly, we represent the average angular velocity by
\[
\dot{\eta} = \frac{1}{2}(\dot{\theta}_1 + \dot{\theta}_2) = \frac{1}{2}(\omega_1 + \omega_2).
\]
The angle difference between each SM and the rotating reference frame is defined by:
\begin{equation}
\delta := \theta_2 - \eta = -(\theta_1 - \eta) = \frac{1}{2}(\theta_2 - \theta_1),
\label{relativeangle}
\end{equation}
where $\delta$ denotes the relative angle between the two generators with respect to the reference frame (\ref{transangleTSGTRL}).
\noindent
To derive the system equations (\ref{TSG_CT_SL3phasemodel}) in the \textit{dq} frame, we consider the transformation angle \(\eta =  \frac{1}{2}(\theta_1 + \theta_2)\), such that  \(i_{abc} = U^\top(\eta) i_{dq}\). The mechanical equation in (\ref{TSG_CT_SL3phasemodel}) is rewritten by substituting \(i_{abc}\), and the electrical equation is multiplied by \(U(\eta)\). After substituting and simplifying, and using the definition  in (\ref{relativeangle}), the \(dq\)-frame representation of the interconnected SMs follows.

 The mechanical subsystem in this model is defined by the equations
\begin{equation}
\begin{aligned}
\dot{\delta} &= \tfrac{1}{2}(\omega_2 - \omega_1), \\[4pt]
J \dot{\omega}_1 &= -D \omega_1 + b(-\sin\delta\, i_{d_1} + \cos\delta\, i_{q_1}) + T_{m_1}, \\[4pt]
J \dot{\omega}_2 &= -D \omega_2 + b(\sin\delta\, i_{d_2} - \cos\delta\, i_{q_2}) + T_{m_2}.
\end{aligned}
\label{mech_block}
\end{equation}
The electrical subsystem in this model is defined by the equations
\begin{equation}
\begin{aligned}
L \dot{i}_{d_1} &= -R i_{d_1} - L\omega i_{q_1} - b\sin\delta \,\omega_1 + R_L i_{d_3}, \\[4pt]
L \dot{i}_{q_1} &= -R i_{q_1} + L\omega i_{d_1} + b\cos\delta \,\omega_1 + R_L i_{q_3}, \\[6pt]
L \dot{i}_{d_2} &= -R i_{d_2} - L\omega i_{q_2} + b\sin\delta \,\omega_2 - R_L i_{d_3}, \\[4pt]
L \dot{i}_{q_2} &= -R i_{q_2} + L\omega i_{d_2} - b\cos\delta \,\omega_2 - R_L i_{q_3}, \\[6pt]
L_3 \dot{i}_{d_3} &= -2R_L i_{d_3} + R_L(i_{d_1}-i_{d_2}) - L_3\omega i_{q_3}, \\[4pt]
L_3 \dot{i}_{q_3} &= -2R_L i_{q_3} + R_L(i_{q_1}-i_{q_2}) + L_3\omega i_{d_3}.
\end{aligned}
\label{elec_block}
\end{equation}

In equations (\ref{elec_block}), $\omega = \tfrac{1}{2}(\omega_1+\omega_2)$ is the average angular velocity.   The mechanical states $(\delta, \omega_1, \omega_2)$ are directly coupled to the electrical currents through the trigonometric terms involving $\delta$. The cross terms $(\pm L\omega)$ capture the $dq$ cross-coupling, while the resistive interactions appear through the symmetric coupling $R,R_L$. The parameter $b$ is defined by $ b = \sqrt{\frac{3}{2}}\, M_f i_f,$ where $M_f$ is the mutual inductance and $i_f$ is the rotor current.

\section{Equilibrium Points for Synchronous Machine Systems}
\label{sec:equilibrium}
\noindent
This section derives the equilibrium points for the two configurations introduced above. For each circuit, the equilibrium is obtained by setting the time derivatives in the corresponding system equations to zero and then solving the resulting algebraic equations.

\subsection{Synchronous generator connected to an impedance load}

\noindent
Here we set the time derivatives in (\ref{electromechanical_modeling_dq}) to zero; which yields
\begin{equation}
\begin{aligned}
    0 &= -D \omega^e - b i_q^e + T_m,  \\
    0 &= - R i_d^e - L\omega^e i_q^e, \\
    0 &= L\omega^e i_{d}^e - R i_q^e + b \omega{^e}.
\end{aligned}
\label{eqfist}
\end{equation}
\noindent
From the second equation in (\ref{eqfist}), we obtain
\begin{equation}
   i_d^e = -\frac{L\omega^e}{R} i_q^e.
\label{ide_SGRL}
\end{equation}
Substituting into the third equation of (\ref{eqfist}) yields
\begin{equation}
i_q^e = \frac{b \omega^e}{\frac{L^2 \omega{^e}^2}{R} + R}.
\label{iqe_SGRL}
\end{equation}
Substituting \(i_q^e\) into the first equation of (\ref{eqfist}) yields the following polynomial equation in \(\omega^e\):
\begin{equation}
DL^2 \omega{^e}^3 - T_m L^2 \omega{^e}^2 + (b^2 R^2 + DR^2)\omega{^e} - T_m R^2 = 0.
\label{omegaequation}
\end{equation}
\noindent
The signs of the coefficients in \eqref{omegaequation} alternate in the pattern (+, –, +, –), which, according to Descartes' rule of signs, implies that the equation admits either one or three positive real roots. Since all coefficients are real and the leading coefficient is positive while the constant term is negative, the existence of at least one positive real root is guaranteed. To express the equation in standard cubic form, we divide by \(DL^2\) and rewrite (\ref{omegaequation}) as
\[
\omega{^e}^3 - \frac{T_m}{D} \omega{^e}^2 + \frac{R^2(b^2 + D)}{DL^2} \omega{^e} - \frac{T_m R^2}{DL^2} = 0.
\]
\noindent
We let \(\omega{^e} = y + \frac{T_m}{3D}\), and eliminating the quadratic term, we obtain
\[
y^3 + p y + q = 0,
\]
where
\[
p = -\frac{T_m^2}{3D^2} + \frac{R^2(b^2 + D)}{DL^2}, \quad
q = -\frac{2T_m^3}{27D^3} + \frac{T_m R^2(b^2 - 2D)}{3D^2L^2}.
\]
\noindent
Using Cardano’s formula \cite{tokunaga1991triple}, the real root of the original cubic equation  (\ref{omegaequation}) is
\begin{equation}
\omega{^e} = \frac{T_m}{3D} + \sqrt[3]{-\frac{q}{2} + \sqrt{\Delta}} + \sqrt[3]{-\frac{q}{2} - \sqrt{\Delta}}.
\label{rootdelta}
\end{equation}
Here
\[
\Delta = \left(\frac{q}{2}\right)^2 + \left(\frac{p}{3}\right)^3.
\]
\noindent
Expanding the formula for \( \Delta \), we obtain a rational expression as follows:
{\small
\begin{equation}
  \Delta =\frac{4 T_m^4 R^2 L^4+T_m^2 R^4L^2 (8 D^2-b^4 - 20 b^2 D)+ 4R^6 D (b^2+ D)^3}{108 D^4 L^6}.
\label{Delta1cir}  
\end{equation}
}
\noindent
The nature of the roots of \eqref{omegaequation} depends on the sign of \(\Delta\) as follows:
\begin{enumerate}
    \renewcommand{\labelenumi}{(\roman{enumi})}
    \item \(\Delta > 0\): one real root, two complex roots;
    \item \(\Delta = 0\): all roots real, with at least two equal;
    \item \(\Delta < 0\): three distinct real roots.
\end{enumerate}

\subsection{Two interconnected SMs supplying resistive loads}
\noindent
To calculate the equilibrium points in this case, we obtain the following set of equations:

\begin{subequations}\label{eq:nonlinearmodel}
    \begin{align}
\omega_1 = \omega_2=\omega^e ,\label{e1} \\
-D \omega^e + b \sin(\delta^e) i_{d_1}^e - b \cos(\delta^e) i_{q_1}^e + T_{m_1} = 0,\label{e2} 
\\
-D \omega^e - b \sin(\delta^e) i_{d_2}^e - b \cos(\delta^e) i_{q_2}^e + T_{m_2} = 0,\label{e3} \\
-b \sin(\delta^e) \omega^e - R i_{d_1}^e - L\omega^e i_{q_1}^e + R_L i_{d_3}^e = 0, \label{e4}\\
b \sin(\delta^e) \omega^e - R i_{d_2}^e - L\omega^e i_{q_2}^e - R_L i_{d_3}^e = 0,\label{e5} \\
b \cos(\delta^e) \omega^e + L\omega^e i_{d_1}^e - R i_{q_1}^e + R_L i_{q_3}^e = 0, \label{e6}\\
b \cos(\delta^e) \omega^e + L\omega^e i_{d_2}^e - R i_{q_2}^e - R_L i_{q_3}^e = 0,\label{e7} \\
R_L i_{d_1}^e - R_L i_{d_2}^e - 2R_L i_{d_3}^e - L_3\omega^e i_{q_3}^e = 0,\label{e8}\\
R_L i_{q_1}^e - R_L i_{q_2}^e + L_3\omega^e i_{d_3}^e - 2R_L i_{q_3}^e = 0,\label{e9}
    \end{align}
\end{subequations}

% \[
% \begin{aligned}
% \omega_1 = \omega_2=\omega^e \quad\quad \text{(19.1)}\\
% -D \omega^e + b \sin(\delta^e) i_{d_1}^e - b \cos(\delta^e) i_{q_1}^e + T_{m_1} = 0, \quad\quad \text{(19.2)}
% \\
% -D \omega^e - b \sin(\delta^e) i_{d_2}^e - b \cos(\delta^e) i_{q_2}^e + T_{m_2} = 0, \quad\quad\text{(19.3)}\\
% \\
% -b \sin(\delta^e) \omega^e - R i_{d_1}^e - L\omega^e i_{q_1}^e + R_L i_{d_3}^e = 0, \quad\quad\text{(19.4)}\\
% b \sin(\delta^e) \omega^e - R i_{d_2}^e - L\omega^e i_{q_2}^e - R_L i_{d_3}^e = 0,\quad\quad\text{(19.6)} \\
% \\
% b \cos(\delta^e) \omega^e + L\omega^e i_{d_1}^e - R i_{q_1}^e + R_L i_{q_3}^e = 0,\quad\quad\text{(19.5)} \\
% b \cos(\delta^e) \omega^e + L\omega^e i_{d_2}^e - R i_{q_2}^e - R_L i_{q_3}^e = 0, \quad\quad\text{(19.7)}\\
% \\
% R_L i_{d_1}^e - R_L i_{d_2}^e - 2R_L i_{d_3}^e - L_3\omega^e i_{q_3}^e = 0,\quad\quad \text{(19.8)}\\
% R_L i_{q_1}^e - R_L i_{q_2}^e + L_3\omega^e i_{d_3}^e - 2R_L i_{q_3}^e = 0,
% \quad\quad \text{(19.9)}
% \end{aligned}
% \]
Rearranging  \eqref{e8} and \eqref{e9}, we obtain
\[
\begin{aligned}
i_{d_3}^e &= \frac{2 R_L^2 (i_{d_1}^e - i_{d_2}^e) - L_3\omega^e R_L (i_{q_1}^e - i_{q_2}^e)}{4 R_L^2 + (L_3\omega^e)^2}, \\
i_{q_3}^e &= \frac{L_3\omega^e R_L (i_{d_1}^e - i_{d_2}^e) + 2 R_L^2 (i_{q_1}^e - i_{q_2}^e)}{4 R_L^2 + (L_3\omega^e)^2}.
\end{aligned}
\]
Next, we substitute \( i_{d_3}^e, i_{q_3}^e \) into equations \eqref{e4}–\eqref{e7} to obtain
\[
X = 4R_L^2 + (L_3 \omega^e)^2.
\]  
We now express these equations in matrix form
\[
\begin{bmatrix}
A & B \\
B & A
\end{bmatrix}
\begin{bmatrix}
i_1 \\
i_2
\end{bmatrix}
=
\begin{bmatrix}
s_1 \\
s_2
\end{bmatrix},
\]
where, the sub-matrices \(A\) and \(B\) are given by
\[\begin{aligned}
A&=\begin{bmatrix}
 -R+\dfrac{2R_L^3}{X} & -L\omega^e-\dfrac{L_3\omega^e R_L^2}{X} \\
 L\omega^e+\dfrac{L_3\omega^e R_L^2}{X} & -R+\dfrac{2R_L^3}{X}
 \end{bmatrix},\\
B&=\begin{bmatrix}
-\dfrac{2R_L^3}{X}  & \dfrac{L_3\omega^e R_L^2}{X} \\
-\dfrac{L_3\omega^e R_L^2}{X} & -\dfrac{2R_L^3}{X}
\end{bmatrix}.
\end{aligned}
\]
Also, the current and source vectors are defined by
\[
i_{j\in \{1, 2\}}^e = \begin{bmatrix} i_{d_{j}}^e \\ i_{q_{j}}^e \end{bmatrix},
s_1 = \begin{bmatrix} b \sin(\delta^e)\omega^e \\ -b \cos(\delta^e)\omega^e \end{bmatrix},
s_2 = \begin{bmatrix} -b \sin(\delta^e)\omega^e \\ -b \cos(\delta^e)\omega^e \end{bmatrix}.
\]
To decouple the equations, we define the sum and difference variables  
\[
v = i_1 + i_2, \quad w = i_1 - i_2,
\]  
which provides the relationships  
\[
(A + B)v = s_1 + s_2, \quad (A - B)w = s_1 - s_2.
\]  
Accordingly,  
\[
v = (A + B)^{-1}(s_1 + s_2), \quad
w = (A - B)^{-1}(s_1 - s_2),
\]
and the original currents are recovered as  
\[
i_1 = \tfrac{1}{2}(v + w), \quad i_2 = \tfrac{1}{2}(v - w).
\]

After substituting and simplifying, the currents can be expressed in closed form using the parameters  
\[
a = -R + \dfrac{4R_L^3}{X}, \quad 
e = -L\omega^e - \dfrac{2L_3 \omega^e R_L^2}{X}, \quad 
R = R_s + R_L,
\]  
as
\[
\begin{aligned}
i_{d_1}^e &= \dfrac{b \omega^e}{2} \left( \frac{-L \omega^e \cos \delta^e}{R^2 + L^2 {\omega^e}^2} + \frac{a \sin \delta^e}{a^2 + e^2} \right), \\
i_{q_1}^e &= \dfrac{b \omega^e}{2} \left( \frac{R \cos \delta^e}{R^2 + L^2 {\omega^e}^2} - \frac{e \sin \delta^e}{a^2 + e^2} \right), \\[6pt]
i_{d_2}^e &= \dfrac{b \omega^e}{2} \left( \frac{-L \omega^e \cos \delta^e}{R^2 + L^2 {\omega^e}^2} - \frac{a \sin \delta^e}{a^2 + e^2} \right), \\
i_{q_2}^e &= \dfrac{b \omega^e}{2} \left( \frac{R \cos \delta^e}{R^2 + L^2 {\omega^e}^2} + \frac{e \sin \delta^e}{a^2 + e^2} \right).
\end{aligned}
\]
By substituting the expressions for the currents into the summation and difference of equations \eqref{e2}–\eqref{e3}, the following relationships are obtained. From the sum, we obtain  
\[
 -2D\omega^e 
 + b^2 \omega^e \left( \frac{a \sin^2 \delta^e}{a^2 + e^2} 
 - \frac{R \cos^2 \delta^e}{R^2 + L^2 {\omega^e}^2} \right) 
 + T_{m_1} + T_{m_2} = 0,
\]  
and from the difference, we obtain 
\[
 b^2 \omega^e \left( \frac{-L \omega^e \sin\delta^e \cos\delta^e}{R^2 + L^2 {\omega^e}^2} 
 + \frac{e \sin\delta^e \cos\delta^e}{a^2 + e^2} \right) 
 + T_{m_1} - T_{m_2} = 0.
\]

From the difference, we can explicitly solve for \(\sin\delta^e \cos\delta^e\) as
\begin{equation}
\begin{aligned}
\sin\delta^e \cos\delta^e &= \dfrac{T_{m_2} - T_{m_1}}{b^2 \omega^e \left( \dfrac{-L\omega^e}{R^2 + L^2 {\omega^e}^2} + \dfrac{e}{a^2 + e^2} \right)}.
% \sin(2\delta^e) &= \dfrac{2(T_{m_2} - T_{m_1})}{b^2 \omega^e \left( \dfrac{-L\omega^e}{R^2 + L^2 {\omega^e}^2} + \dfrac{e}{a^2 + e^2} \right)}.
\end{aligned}
\label{sin}
\end{equation}

From the sum and using the trigonometric identity \(\sin^2\delta^e + \cos^2\delta^e = 1\), and defining  
\[
T_d = T_{m_2} - T_{m_1}, \quad T_s = T_{m_2} + T_{m_1},
\]  
we obtain  
\begin{equation}
\sin^2\delta^e = \frac{ \left( 2D\omega^e - T_s \right)(R^2 + L^2 {\omega^e}^2) + R b^2 \omega^e }{ \left( a(R^2 + L^2 {\omega^e}^2) + R(a^2 + e^2) \right) } \cdot \frac{a^2 + e^2}{b^2 \omega^e},
\label{sin2}
\end{equation}  
\begin{equation}
\cos^2\delta^e = \frac{ a b^2 \omega^e + (T_s - 2D\omega^e)(a^2 + e^2) }{ \left( a(R^2 + L^2 {\omega^e}^2) + R(a^2 + e^2) \right) } \cdot \frac{R^2 + L^2 {\omega^e}^2}{b^2 \omega^e}.
\label{cos2}
\end{equation}

To obtain a polynomial in terms of $\omega^e$,  we use the equality  $(\sin\delta^e \cos\delta^e)^2=\cos^2\delta^e \sin^2\delta^e $.
Then, simplifying and rearranging this equality we obtain
 % canceling the common factors and making rearranging the equality, we obtain:
\[
\begin{aligned}
&\Big((2D\omega^e - T_s)(R^2 + L^2{\omega^e}^2) + R b^2 \omega^e\Big) \nonumber \\
&\quad \times \Big[(-R(4R_L^2 + L_3^2 {\omega^e}^2) + 4R_L^3) b^2 \omega^e \nonumber \\
&\qquad + (T_s - 2D\omega^e)\Big((R^2 + L^2{\omega^e}^2)(4R_L^2 + L_3^2 {\omega^e}^2) \nonumber \\
&\qquad \quad + R_L^2 (4L L_3 {\omega^e}^2 - 8R R_L) + 4 R_L^4\Big)\Big] \nonumber \\
&\quad \times \Big[-2L\omega^e (R^2 + L^2{\omega^e}^2)(4R_L^2 + L_3^2 {\omega^e}^2) \nonumber \\
&\qquad - 2L_3 \omega^e R_L^2 (R^2 + L^2{\omega^e}^2) \nonumber \\
&\qquad - L\omega^e \Big(R_L^2 (4L L_3 {\omega^e}^2 - 8R R_L) + 4 R_L^4\Big)\Big]^2 \nonumber \\
&= T_d^2 (R^2 + L^2{\omega^e}^2) \Big[(R^2 + L^2{\omega^e}^2)(4R_L^2 + L_3^2 {\omega^e}^2) \nonumber \\
&\quad + R_L^2 (4L L_3 {\omega^e}^2 - 8R R_L) + 4 R_L^4\Big] \times \nonumber \\
& \Big[4R_L^3 (R^2 + L^2{\omega^e}^2) + R\Big(R_L^2 (4L L_3 {\omega^e}^2 - 8R R_L) + 4 R_L^4\Big)\Big]^2.
\end{aligned}
\]
The resulting polynomial is of order 18 with coefficients that depend on the system parameters. Due to the complicated form, the full expression is provided in the appendix, and a simplified version is as follows:
\begin{equation}
\begin{aligned}
&- H_0 + (K_0 G_2 - H_2) {\omega^e}^2 + K_1 G_2 {\omega^e}^3+ \\
& (K_0 G_4 + K_2 G_2 - H_4) {\omega^e}^4 + (K_1 G_4 + K_3 G_2) {\omega^e}^5+ \\
& (K_0 G_6 + K_2 G_4 + K_4 G_2 - H_6) {\omega^e}^6\\
& + (K_1 G_6 + K_3 G_4 + K_5 G_2) {\omega^e}^7 \\
&+(K_0 G_8 + K_2 G_6 + K_4 G_4 + K_6 G_2 - H_8) {\omega^e}^8+ \\
& (K_1 G_8 + K_3 G_6 + K_5 G_4 + K_7 G_2) {\omega^e}^9+ \\
& (K_0 G_{10} + K_2 G_8 + K_4 G_6 + K_6 G_4 + K_8 G_2 - H_{10}) {\omega^e}^{10}+ \\
& (K_1 G_{10} + K_3 G_8 + K_5 G_6 + K_7 G_4) {\omega^e}^{11}+ \\
& (K_2 G_{10} + K_4 G_8 + K_6 G_6 + K_8 G_4) {\omega^e}^{12}+ \\
& (K_3 G_{10} + K_5 G_8 + K_7 G_6) {\omega^e}^{13} \\
& + (K_4 G_{10} + K_6 G_8 + K_8 G_6) {\omega^e}^{14} \\
&+ (K_5 G_{10} + K_7 G_8) {\omega^e}^{15} + (K_6 G_{10} + K_8 G_8) {\omega^e}^{16} \\
&+ K_7 G_{10} {\omega^e}^{17} - K_8 G_{10} {\omega^e}^{18} = 0.
\end{aligned}
\label{TSGTRL_omegapoly}
\end{equation}

An analytical solution of equation~\eqref{TSGTRL_omegapoly} is not possible; therefore, its roots must be determined numerically. For the solutions to be physically meaningful, the system parameters must satisfy the following conditions. \begin{enumerate}
    \renewcommand{\labelenumi}{(\roman{enumi})}
    \item The absolute magnitude of the expression in equation~\eqref{sin} must be less than or equal to 0.5.
    \item The sum of the expressions in equations~\eqref{sin2} and~\eqref{cos2} must equal one.
    \item The solution \(\omega^e\) must be a positive real number.
\end{enumerate}

\section{Stability Analysis for Synchronous Machine Systems}
\label{sec:stability}
This section examines the stability of the equilibrium points for the two synchronous machine systems.  
Specifically, the SG connected to an impedance load and two SMs serving resistive loads.
For the SG connected to an impedance load, Lyapunov’s direct method is used as well as a linearization method. In contrast, for the two synchronous generator system, we consider specific parameter values and then linearize the corresponding equations around the equilibrium points.  

\subsection{SG connected to impedance load}

\noindent Here we use the Lyapunov direct  method, and obtain the perturbed system dynamics by substituting \(\delta=\tilde{\delta}+\delta^e , \omega=\tilde{\omega}+\omega^e, i_d=\tilde{i}_d+i_d^e,i_q= \tilde{i}_q+i_q^e\) as follows
\[
\begin{aligned}
    \dot{\tilde{\delta}}&=\tilde{\omega}, \\
    J\dot{\tilde{\omega}}&=-D (\tilde{\omega}+\omega^e) - b(\tilde{i}_q+i_q^e) + T_m, \\
    L\dot{\tilde{i}}_d&=- R (\tilde{i}_d+i_d^e)- L(\tilde{\omega}+\omega^e) (\tilde{i}_q+i_q^e), \\
    L\dot{\tilde{i}}_q&=b(\tilde{\omega}+\omega^e) + L(\tilde{\omega}+\omega^e) (\tilde{i}_d+i_d^e) - R (\tilde{i}_q+i_q^e).
\end{aligned}
\]
At the equilibrium point, $ -D \omega^e - b i_q^e + T_m=0  ,- R i_d^e - L \omega^e i_q^e=0  , b \omega^e + L \omega^e i_d^e - R i_q^e=0 $, and hence the perturbed system dynamics are:
\begin{equation}
\begin{aligned}
    \dot{\tilde{\delta}}&=\tilde{\omega}, \\
    J\dot{\tilde{\omega}}&=-D \tilde{\omega} - b\tilde{i}_q , \\
    L\dot{\tilde{i}}_d&=- R \tilde{i}_d- L\tilde{\omega} \tilde{i}_q-L\omega^e \tilde{i}_q-L\tilde{\omega} i_q^e, \\
    L\dot{\tilde{i}}_q&=b\tilde{\omega} +  L\tilde{\omega} \tilde{i}_d+L\omega^e \tilde{i}_d+L\tilde{\omega} i_d^e- R \tilde{i}_q.
\end{aligned}
\label{perturbcircuit1}
\end{equation}
A candidate Lyapunov function is constructed as the sum of the kinetic energy and the energy stored in the inductor, L, given by the quadratic form
\[
    V(\tilde{\omega} , \tilde{i}_{d},\tilde{i}_{q}) = \frac{J}{2} \tilde{\omega}^2 + \frac{L}{2}(\tilde{i}_{d}^2 + \tilde{i}_{q}^2).
\]
The derivative of the candidate Lyapunov function can be written as
\[
\dot V =\begin{bmatrix}
    \tilde{\omega} & \tilde{i}_{d} & \tilde{i}_{q}
\end{bmatrix} 
Q
\begin{bmatrix}
    \tilde{\omega} \\ \tilde{i}_{d}  \\ \tilde{i}_{q}
\end{bmatrix} ,  \quad Q = \begin{bmatrix} 
    -D & -\dfrac{Li_q^e}{2} & \dfrac{L i_d^e}{2} \\ 
    -\dfrac{L i_q^e}{2} & -R & 0 \\ 
    \dfrac{L i_d^e}{2} & 0 & -R 
\end{bmatrix}.
\]
To determine the negative semi-definiteness of the matrix \( Q \), we analyze its leading principal minors, noting that \( Q \) is symmetric:

\begin{enumerate}
    \renewcommand{\labelenumi}{(\roman{enumi})}
    \item The first leading principal minor is negative if and only if:  
    \[
    D \geq 0.
    \]
    
    \item The second leading principal minori is positive if and only if:  
    \[
    \begin{vmatrix} 
    -D & -\frac{L i_q^e}{2} \\ 
    -\frac{L i_q^e}{2} & -R 
    \end{vmatrix} = DR - \left(\frac{L i_q^e}{2}\right)^2 \geq 0,
    \]
    which is equivalent to the condition
    \[
    DR \geq \frac{L^2 (i_q^e)^2}{4}.
    \]
    
    \item The third leading principal minor is negative if and only if:  
    \[
     -D R^2 + \frac{L^2 R}{4} \left( (i_q^e)^2 - (i_d^e)^2 \right) \leq 0,
    \]
    \noindent which is equivalent to
    \[
    D R \geq \frac{L^2}{4} \left( (i_d^e)^2 - (i_q^e)^2 \right).
    \]
\end{enumerate}

Combining these conditions, we conclude that the matrix \( Q \) is negative semi-definite if and only if  the following inequality holds:
\begin{equation}
D \geq \frac{L^2}{4R} \left( (i_q^e)^2 + (i_d^e)^2 \right).
\label{conditionfirstc}
\end{equation}
Substituting \( i_d^e \) and \( i_q^e \) from (\ref{ide_SGRL}), (\ref{iqe_SGRL}) into the inequality (\ref{conditionfirstc}) yields the condition
\begin{equation}
D \geq \frac{b^2}{4R} \left(\frac{L^2 {\omega^e}^2}{R^2 + L^2 {\omega^e}^2} \right).   
\label{conditionC1L}
\end{equation}
The inequality ~\eqref{conditionC1L} indicates that an increase in the damping coefficient \(D\) enhances system stability. Moreover, lower values of the \(\omega^e\) contribute to a larger stability margin compared to higher angular velocities. In addition, an increase in the excitation current \(b\) enlarges the right-hand side of the inequality, thereby reducing the overall region of stability. These trends are physically intuitive and align well with practical observations in real-world systems.

% \subsubsection{Linearization for Multiple Equilibrium Points}

Linearizing the system around the equilibrium point \((\omega^e, i_d^e, i_q^e)\) and neglecting second-order terms gives:
\begin{equation}
\begin{aligned}
J\dot{\tilde{\omega}} &= -D\,\tilde{\omega} - b\,\tilde{i}_q, \\
L\dot{\tilde{i}}_d &= - R\,\tilde{i}_d - L\omega^e \,\tilde{i}_q - L i_q^e \,\tilde{\omega}, \\
L\dot{\tilde{i}}_q &= b\,\tilde{\omega} + L\omega^e\,\tilde{i}_d + L i_d^e \,\tilde{\omega} - R\,\tilde{i}_q.
\end{aligned}
\end{equation}
The linearized perturbed system in state space form is as follows:
\[
\begin{bmatrix}
\dot{\tilde{\omega}} \\
\dot{\tilde{i}}_d \\
\dot{\tilde{i}}_q
\end{bmatrix}
=
\begin{bmatrix}
-\frac{D}{J} & 0 & -\frac{b}{J} \\
-\frac{i_q^e}{L} & -\frac{R}{L} & -\omega^e \\
\frac{b + i_d^e}{L} & \omega^e & -\frac{R}{L}
\end{bmatrix}
\begin{bmatrix}
\tilde{\omega} \\
\tilde{i}_d \\
\tilde{i}_q
\end{bmatrix},
\]
where \(i_d^e\) and \(i_q^e\) are given by \eqref{ide_SGRL} and \eqref{iqe_SGRL}, respectively.

To assess the stability of the linearized system, we apply the Routh–Hurwitz criterion to its characteristic polynomial, which can be expressed as:
\[
P(\lambda) = \lambda^3 + a_2 \lambda^2 + a_1 \lambda + a_0,
\]
where
\[
\begin{aligned}
a_2 &= \frac{2R}{L} + \frac{D}{J}, \\
a_1 &= \frac{R^2}{L^2} + (\omega^e)^2 + \frac{2RD}{LJ} + \frac{b^2 R^2}{JL \big(L^2 (\omega^e)^2 + R^2\big)}, \\
a_0 &= \frac{D\big(R^2 + L^2 (\omega^e)^2\big)}{JL^2} + \frac{b^2 R\big(R^2 - L^2 (\omega^e)^2\big)}{JL^2 \big(L^2 (\omega^e)^2 + R^2\big)}.
\end{aligned}
\]
According to the Routh-Hurwitz stability criterion, the system is asymptotically stable if and only if:
\begin{equation}
a_2 > 0, \quad a_1 > 0, \quad a_0 > 0, \quad a_2 a_1 > a_0.
\label{RHcriteria}
\end{equation}

\subsection{Two interconnected synchronous machines serving loads}
The nonlinear system of equations in~\eqref{mech_block}--\eqref{elec_block} is linearized around the equilibrium point
\[
x^e = \big(\delta^e,\,\omega_1^e,\,\omega_2^e,\,i_{d_1}^e,\,i_{q_1}^e,\,i_{d_2}^e,\,i_{q_2}^e,\,i_{d_3}^e,\,i_{q_3}^e\big),
\]
where the steady-state condition \(\dot{x}=0\) holds. Next, we obtain the following linearized state-space model
\[
\Delta \dot{x} = A\,\Delta x + B\,\Delta u,
\]
where \(\Delta x = x - x^e\) and \(\Delta u = u - u^e\). The state and input vectors are defined by
\[
    \begin{aligned}
        x &=
        \begin{bmatrix}
        \delta & \omega_1 & \omega_2 & i_{d_1} & i_{q_1} & i_{d_2} & i_{q_2} & i_{d_3} & i_{q_3}
        \end{bmatrix}^\top,\\
        u &=
        \begin{bmatrix}
         \tfrac{T_{m_1}}{J} & \tfrac{T_{m_2}}{J} 
        \end{bmatrix}^\top.
    \end{aligned}
\]

The corresponding Jacobian matrices are calculated as
\[
A = \left. \frac{\partial f}{\partial x} \right|_{x^e,\,u^e},
\qquad
B = \left. \frac{\partial f}{\partial u} \right|_{x^e,\,u^e},
\]
where \(A\) has the block structure:
\[
A =
\begin{bmatrix}
A_{mm} & A_{me} \\[1ex]
A_{em} & A_{ee}
\end{bmatrix}_{9\times 9},
\]
with the submatrices defined as follows:
\begin{enumerate}
    \renewcommand{\labelenumi}{(\roman{enumi})}
\item {mechanical subsystem  }%subsystem 
\[
A_{mm} =
\begin{bmatrix}
0 & -0.5 & 0.5 \\
\dfrac{b(\cos\delta^e i_{d_1}^e + \sin\delta^e i_{q_1}^e)}{J} & -\dfrac{D}{J} & 0 \\
\dfrac{b(-\cos\delta^e i_{d_2}^e + \sin\delta^e i_{q_2}^e)}{J} & 0 & -\dfrac{D}{J}
\end{bmatrix} ,
\]

\item {mechanical--electrical coupling}
\[
A_{me} =
\begin{bmatrix}
0 & 0 & 0 & 0 & 0 & 0 \\
\dfrac{b \sin\delta^e}{J} & \dfrac{-b \cos\delta^e}{J} & 0 & 0 & 0 & 0 \\
0 & 0 & \dfrac{-b \sin\delta^e}{J} & \dfrac{-b \cos\delta^e}{J} & 0 & 0
\end{bmatrix},
\]

\item{electrical--mechanical coupling}
\[
A_{em} =
\begin{bmatrix}
-\dfrac{b \omega^e \cos\delta^e}{L} & -\dfrac{b\sin\delta^e}{L}-\dfrac{i_{q_1}^e}{2} & -\dfrac{i_{q_1}^e}{2}  \\
-\dfrac{b \omega^e \sin\delta^e}{L} & \dfrac{b \cos\delta^e}{L} + \dfrac{i_{d_1}^e}{2}  & \dfrac{i_{d_1}^e}{2}  \\
\dfrac{b \omega^e \cos\delta^e}{L} & -\dfrac{i_{q_2}^e}{2}  & \dfrac{b \sin\delta^e}{L} - \dfrac{i_{q_2}^e}{2}  \\
-\dfrac{b \omega^e \sin\delta^e}{L} & \dfrac{i_{d_2}^e}{2}  & \dfrac{b \cos\delta^e}{L} + \dfrac{i_{d_2}^e }{2} \\
0 & -\dfrac{ i_{q_3}^e}{2} & -\dfrac{i_{q_3}^e}{2}  \\
0 & \dfrac{ i_{d_3}^e}{2} & \dfrac{i_{d_3}^e}{2} 
\end{bmatrix},
\]

\item{electrical subsystem }
\[
A_{ee} =
\begin{bmatrix}
-\dfrac{R}{L} & - \omega^e & 0 &0 & \dfrac{R_L}{L} & 0  \\
\omega^e & -\dfrac{R}{L} & 0 & 0 & 0 & \dfrac{R_L}{L} \\
0 & 0 & -\dfrac{R}{L} & -\omega^e & -\dfrac{R_L}{L} & 0 \\
0 & 0 &  \omega^e & -\dfrac{R}{L} & 0 & -\dfrac{R_L}{L} \\
\dfrac{R_L}{L_3} & 0 & -\dfrac{R_L}{L_3} & 0 & -2 \dfrac{R_L}{L_3} & - \omega^e \\
0 & \dfrac{R_L}{L_3} & 0 & -\dfrac{R_L}{L_3} &  \omega^e & -\dfrac{2R_L}{L_3}
\end{bmatrix}.
\]
\end{enumerate}
\subsection{Illustrative Examples}
Case 1: Synchronous generator connected to a resistive load

The sign of \(\Delta\) in ~\eqref{Delta1cir} determines the number of equilibrium points. It is possible to obtain one, two, or three real equilibria based on selection of the circuit parameters. For the parameter set
\[
D = L = R = 1,\quad b = 4,\quad T_m = 9,
\]
the cubic equation (\ref{omegaequation}) governing the equilibrium angular velocity reduces to
\[
\omega^{e^3} - 9\omega^{e^2} + 17\omega^e - 9 = 0 
\quad \Rightarrow \quad (\omega^e - 1)(\omega^{e^2} - 8\omega^e + 9)=0,
\]
which yields three distinct positive real roots:
\[
\omega^e = 1,\quad \omega^e = 4 \pm \sqrt{7}.
\]
For these parameter values, the Routh–Hurwitz conditions in~\eqref{RHcriteria} are satisfied at two equilibrium points, \(\omega^e = 1\) and \(\omega^e = 4 + \sqrt{7}\), confirming local stability. In contrast, at the equilibrium point \(\omega^e = 4 - \sqrt{7}\), the condition \(a_0 < 0\) is violated, indicating that this operating point is unstable.

\subsubsection{Remarks}

It is important to note that Lyapunov’s direct method provides only sufficient conditions for stability. Consequently, failure to satisfy the Lyapunov condition does not necessarily imply instability. In this example, while Lyapunov’s method cannot guarantee stability for certain equilibria, the linearized analysis confirms that two operating points are indeed locally stable. 

In a numerical simulation with the initial condition \(\omega_0 = 4.5\), although this value lies closer to the equilibrium \(\omega^e = 4 + \sqrt{7}\), the trajectory converges to the equilibrium point \(\omega^e = 1\).

% \begin{figure}[h]
%     \centering
%     \includegraphics[width=1\linewidth]{Pic/5.png}
%     \caption{Phase trajectory convergence for Circuit~1.}
%     \label{circuit1_ex}
% \end{figure}
Case 2: Two synchronous machines serving loads

 We select the parameter set
\[
    \begin{aligned}
   &J = 1, \quad R = 1010, \quad L = 0.041, \quad D = 9, \quad  T_{m1} = 2910, \\
    &T_{m2} = 2800, \quad b = 5, \quad R_L = 1000, \quad L_3 = 0.04. 
    \end{aligned}
\]
In this case, the system admits 18 equilibrium points. Among them, only solutions
\[
\omega^e_1 = 315.5902 \ \mathrm{rad/s}, 
\qquad 
\omega^e_2 = 309.0166 \ \mathrm{rad/s},
\]
satisfy the positive-real condition together with conditions~\eqref{sin}, \eqref{sin2}, and \eqref{cos2}.  
 It is found that the equilibrium point \(\omega^e_1\) is locally stable, as all eigenvalues of the Jacobian have negative real parts, while \(\omega^e_2\) is not locally stable.

These examples highlight the crucial role of grid parameters can play in shaping power system behavior. Even in relatively simple configurations, multiple equilibria can arise. In practical scenarios, the coexistence of nearby equilibria—particularly if more than one is stable—may cause operational challenges. Such situations can make the system sensitive to perturbations, potentially driving the dynamics toward an unintended equilibrium point.

\section{Conclusion }
\label{sec:conclusion}
This paper has presented a detailed study of equilibrium points and stability for two  synchronous machine systems: (1) a single generator with an impedance load, and (2) two interconnected machines with resistive loads. By formulating the models in both $abc$ and $dq$ reference frames, we showed that the equilibrium conditions reduce to a cubic polynomial in the single-machine case and to an 18th order polynomial in the two-machine case. For the single-machine system, Lyapunov analysis and a linearization analysis were carried out. For the two-machine system, where nonlinearities hinder the construction of an energy-based Lyapunov function, stability was investigated through linearization and eigenvalue analysis. Numerical simulations further confirmed the analytical findings and illustrated the dependence of equilibria and stability on system parameters.
\appendix

\section{coefficients of equation (\ref{TSGTRL_omegapoly})}
\vspace{-\abovedisplayskip}
The following are formulas for the coefficients of the polynomial (\ref{TSGTRL_omegapoly}) in terms of the system parameters:

\vspace{-\abovedisplayskip}
\begin{align*}
&{K_0}=  -4 R_L^4 R^2 T_s^2 + 8 R_L^3 R^3 T_s^2 - 4 R_L^2 R^4 T_s^2, \\
&{K_1}=  4 b^2 R_L^4 R T_s - 12 b^2 R_L^3 R^2 T_s + 16 D R_L^4 R^2 T_s\\ &\quad + 8 b^2 R_L^2 R^3 T_s - 32 D R_L^3 R^3 T_s + 16 D R_L^2 R^4 T_s,\\
& {K_2}=  4 b^4 R_L^3 R - 8 b^2 D R_L^4 R - 4 b^4 R_L^2 R^2 + 24 b^2 D R_L^3 R^2\\
&\quad- 16 D^2 R_L^4 R^2 - 16 b^2 D R_L^2 R^3 + 32 D^2 R_L^3 R^3 \\
&\quad - 16 D^2 R_L^2 R^4  - 4 L^2 R_L^4 T_s^2 + 8 L^2 R_L^3 R T_s^2  \\
&\quad - 4 L_3 L R_L^2 R^2 T_s^2 - 8 L^2 R_L^2 R^2 T_s^2 + R^4 T_s^2 \\
&{K_3}=  -4 b^2 L^2 R_L^3 T_s + 16 D L^2 R_L^4 T_s+ 4 L_3 b^2 L R_L^2 R T_s  \\
&\quad + 8 b^2 L^2 R_L^2 R T_s - 32 D L^2 R_L^3 R T_s + 16 L_3 D L R_L^2 R^2 T_s\\
&\quad  + 32 D L^2 R_L^2 R^2 T_s - 2 b^2 R^3 T_s - 4 D R^4 T_s, \\
&{K_4}= 8 b^2 D L^2 R_L^3 - 16 D^2 L^2 R_L^4 - 8 L_3 b^2 D L R_L^2 R\\
&\quad - 16 b^2 D L^2 R_L^2 R + 32 D^2 L^2 R_L^3 R + b^4 R^2 \\
&\quad - 16 L_3 D^2 L R_L^2 R^2 - 32 D^2 L^2 R_L^2 R^2+ 4 b^2 D R^3 \\
&\quad  + 4 D^2 R^4 - 4 L_3 L^3 R_L^2 T_s^2 - 4 L^4 R_L^2 T_s^2 + 2 L^2 R^2 T_s^2, \\
&{K_5}= 16 L_3 D L^3 R_L^2 T_s + 16 D L^4 R_L^2 T_s \\
&\quad - 2 b^2 L^2 R T_s - 8 D L^2 R^2 T_s, \\
&{K_6} =-16 L_3 D^2 L^3 R_L^2 - 16 D^2 L^4 R_L^2\\
&\quad + 4 b^2 D L^2 R + 8 D^2 L^2 R^2 + L^4 T_s^2,\\
&{K_7} =  -4 D L^4 T_s, \\
&{K_8}= 4 D^2 L^4.\\
&{G_{10}} = 4 L^6 , \\
&{G_8}=-32 L^6 R_L^2  - 24 L_3 L^5 R_L^2  + 8 L^4 R^2, \\
&{G_6}=64 L^6 R_L^4 + 96 L_3 L^5 R_L^4 - 52 L^4 R_L^4 + 32 L^4 R_L^3 R\\
&\quad   - 64 L^4 R_L^2 R^2 - 32 L_3 L^3 R_L^2 R^2 + 4 L^2 R^4,\\
&{G_4}=- 8 L_3 L R_L^2 R^4 +64 L^4 R_L^6 - 128 L^4 R_L^5 R+ 128 L^4 R_L^4 R^2  \\
&\quad + 48 L_3 L^3 R_L^6  - 96 L_3 L^3 R_L^5 R+ 128 L_3 L^3 R_L^4 R^2 - 40 L^2 R_L^4 R^2 \\
&\quad  + 32 L^2 R_L^3 R^3  - 32 L^2 R_L^2 R^4, \\
&{G_2}=16 L^2 R_L^8 - 64 L^2 R_L^7 R+ 128 L^2 R_L^6 R^2 - 128 L^2 R_L^5 R^3 \\
&\quad  + 64 L^2 R_L^4 R^4 + 16 L_3 L R_L^6 R^2 - 32 L_3 L R_L^5 R^3 + 32 L_3 L R_L^4 R^4\\
&\quad  - 4 R_L^4 R^4.\\
& {H_{10}}=-16 L^8 T_d^2 R_L^6 - 32 L_3 L^7 T_d^2 R R_L^5 + 16 L^6 T_d^2 R^2 R_L^4, \\[0.5em]
&{H_{8}}=64 L^8 T_d^2 R_L^8 + 64 L_3 L^7 T_d^2 R_L^8 - 160 L^6 T_d^2 R R_L^7\\ &\quad + 128 L_3 L^7 T_d^2 R R_L^7 - 64 L^6 T_d^2 R^2 R_L^6 - 96 L_3 L^5 T_d^2 R^2 R_L^6  \\ &\quad-32 L_3 L^5 T_d^2 R^3 R_L^5 + 32 L^4 T_d^2 R^4 R_L^4, \\[0.5em]
& {H_{6} } = 64 L^6 T_d^2  R_L^{10} - 208 L^4 T_d^2 R^2  R_L^8 - 192 L_3 L^5 T_d^2 R^2  R_L^8 \\ &\quad+ 256 L_3 L^5 T_d^2 R  R_L^9 
 + 128 L_3 L^5 T_d^2 R^3  R_L^7 + 96 L^4 T_d^2 R^3  R_L^7 \\ &\quad - 96 L^4 T_d^2 R^4  R_L^6 - 128 L_3 L^3 T_d^2 R^4  R_L^6  \\ &\quad + 32 L_3 L^3 T_d^2 R^5  R_L^5 + 16 L^2 T_d^2 R^6  R_L^4, \\[0.5em]
& {H_{4}} =128 L^4 T_d^2 R  R_L^{11} - 256 L^4 T_d^2 R^2  R_L^{10} + 192 L_3 L^3 T_d^2 R^2  R_L^{10} \\ &\quad + 256 L^4 T_d^2 R^3  R_L^9  - 256 L_3 L^3 T_d^2 R^3  R_L^9 - 128 L^4 T_d^2 R^4  R_L^8\\ &\quad
+ 192 L_3 L^3 T_d^2 R^4  R_L^8 - 224 L^2 T_d^2 R^4  R_L^8- 128 L_3 L^3 T_d^2 R^5  R_L^7  \\ &\quad + 288 L^2 T_d^2 R^5  R_L^7 - 64 L^2 T_d^2 R^6  R_L^6 - 32 L_3 L T_d^2 R^6  R_L^6  \\ &\quad+ 32 L_3 L T_d^2 R^7  R_L^5,\\
  \end{align*}
 \begin{align*}
& {H_2} =64 L^2 T_d^2 R^2  R_L^{12} - 128 L^2 T_d^2 R^3  R_L^{11} + 64 L^2 T_d^2 R^4  R_L^{10} \\ &\quad+ 192 L_3 L T_d^2 R^4  R_L^{10} - 512 L_3 L T_d^2 R^5  R_L^9  
 +448 L_3 L T_d^2 R^6  R_L^8 \\ &\quad- 16 T_d^2 R^6 R_L^8 - 128 L_3 L T_d^2 R^7  R_L^7  + 32 T_d^2 R^7  R_L^7 - 16 T_d^2 R^8  R_L^6,\\[0.5em]
& {H_0}= 64 T_d^2 R^4 R_L^{12} - 256 T_d^2 R^5 R_L^{11} + 384 T_d^2 R^6 R_L^{10}\\ &\quad - 256 T_d^2 R^7 R_L^9 + 64 T_d^2 R^8 R_L^8.
\end{align*}

\bibliography{ifacconf}           
\end{document}